\documentclass[RNAAS]{aastex63}

\begin{document}
\title{Cloudy in the microcalorimeter era: improved energies for K$\alpha$ transitions }

\author[0000-0002-4469-2518]{P. Chakraborty}
\affiliation{University of Kentucky \\
Lexington, KY, USA}
\author[0000-0003-4503-6333]{G. J. Ferland}
\affiliation{University of Kentucky \\
Lexington, KY, USA}
\author[0000-0002-4622-4240]{S.Bianchi}
\affiliation{Dipartimento di Matematica e Fisica, Universit\`a degli Studi Roma Tre, via della Vasca Navale 84, I-00146 Roma, Italy \\
}
\author[0000-0002-8823-0606]{M. Chatzikos}
\affiliation{University of Kentucky \\
Lexington, KY, USA}




\begin{abstract}
    
X-ray missions with microcalorimeter technology will resolve spectral features with unprecedented detail.
In this work, we improve the H-like K$\alpha$ energies for elements between 6 $\leq Z \leq$ 30 for the release version
of the spectral simulation code Cloudy
to match laboratory energies.
We update the ionization potential ($I_{\rm ion}$) for these elements 
and add a fourth-order polynomial to the level energy difference.  This brings the release version of Cloudy
into a near-perfect agreement with NIST. The updated energies are $\sim$ 15-4000 times more precise than that of
the current release version of Cloudy (C17.02). These new changes will be a part of the next update
to the release version, C17.03.
\end{abstract}

\section{Introduction}

Cloudy \citep{2017RMxAA..53..385F} has long predicted intensities of X-ray lines
due to its need to do physical simulations of a non-equilibrium plasma.
The original design for one and two-electron species 
took advantage of scaling relationships along iso-electronic sequences.
A series of papers, part of Ryan Porter’s thesis, reporting on this development include 
\citet{2005ApJ...622L..73P} and \citet{2012MNRAS.425L..28P} on optical emission from He I and \citet{2007ApJ...664..586P}
on X-ray emission from O VII.
While the physics remains close to state of the art, the level energies and line wavelengths 
derived in the older work
had largely sufficient accuracy for the then-operational optical observatories and X-ray missions but not future missions.

The available spectroscopic resolution has increased dramatically with the advent of 
microcalorimeter missions 
like Hitomi and the upcoming missions XRISM and Athena. 
We are now extending Cloudy to meet
the spectroscopic challenges of such missions as part of Priyanka Chakraborty's thesis.
The first papers focused on two-electron Fe K$\alpha$ emission.
\citet{2020ApJ...901...68C} discussed line interlocking and Resonant Auger Destruction \citep{1978ApJ...219..292R, 1990ApJ...362...90B, 1996MNRAS.278.1082R,2005AIPC..774...99L}, 
and electron scattering escape (ESE) in the Fe XXV K$\alpha$ complex.
\citet{2020ApJ...901...69C} discussed the Case A to B transition in H- and He- like iron.
These atomic processes are very sensitive to line wavelengths due to line overlap with
nearby satellites \citep{2015A&A...579A..87M, 2020ApJ...901...68C} and require precise energies.
This development is a work in progress and will be part of a future release of Cloudy.
In the meantime, we have improved the treatment of levels and line energies
in the release version of Cloudy, as described below. 
These improvements will be part of the C17.03 release in late 2020.

\section{Results}

The previous versions of Cloudy, through to C17.02, used  ionization potentials ($I_{\rm ion}$)
derived by \citet{1996ApJ...465..487V} with
four significant digits. 
Level energies were then derived from the following equation: 
\begin{equation}\label{e:1}
I_n = I_{\rm ion}/ n^{2} 
\end{equation}
and the K$\alpha$ energies were calculated from:

\begin{equation}
E_{\rm K\alpha}^{\rm old} = I_1 - I_2
\end{equation}

The $I_{\rm ion}$'s in equation \ref{e:1}, stored in the `phfit.dat' file in the Cloudy data directory,  
were  used to compute 
the photoionization cross-sections in \citet{1996ApJ...465..487V}. 
Although these values are reasonably accurate,
 microcalorimeter observations  require much better precision. 
We update the $I_{\rm ion}$'s in `phfit.dat' for H-like ions with those of NIST \citep{2018APS..DMPM01004K}, 
keeping up to the eighth significant digits. 
Our updated version of `phfit.dat' will be included in the C17.03 release.

Following this, 
we generated a fourth-order polynomial for the energy correction ($\Delta E$ ) for better agreement with the NIST K$\alpha$ energies:

\begin{equation}\label{e:2}
\Delta E = 0.1783 Z^{4} - 1.8313 Z^{3} +27.803 Z^{2}- 208.04 Z+ 570.59
\end{equation}

The updated energies ($E_{\rm K\alpha}^{\rm new}$) for the K$\alpha$ transitions are given by the following equation:

\begin{equation}
E_{\rm K\alpha}^{\rm new} = E_{\rm K\alpha}^{\rm old} +  \Delta E
\end{equation}
\begin{figure}[h!]
\begin{center}
\includegraphics[scale=0.6,angle=0]{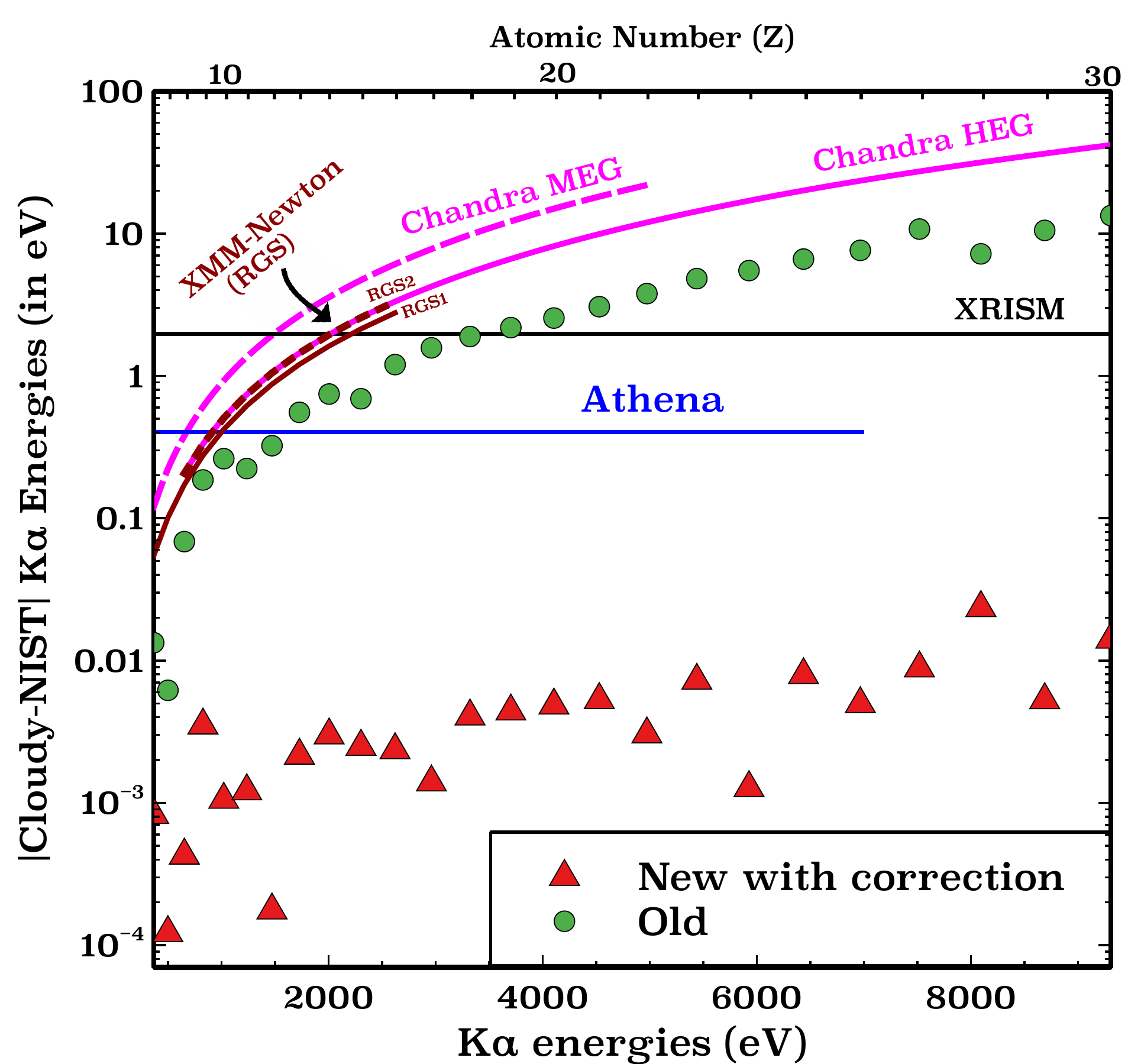}
\caption{The absolute value of the difference between NIST and Cloudy K$\alpha$ energies versus 
K$\alpha$ energies for H-like ions of
elements between $6 \leq Z \leq 30$. The x-axis on top  shows 
the corresponding atomic numbers (Z). Red triangles show the 
difference between energies
in the updated version of Cloudy (C17.03) and NIST. Green circles show 
the energy difference
between the previous versions of Cloudy (from $\sim$2005 to C17.02) 
and NIST. The solid, dashed, and dotted lines indicate the
absolute values of the energy accuracy of the current 
and future X-ray observatories. Refer to the results section for a detailed description. }
\label{fig:1}
\end{center}
\end{figure}

Figure \ref{fig:1} shows the absolute values of the differences in K$\alpha$ energies between NIST and Cloudy for H-like ions versus K$\alpha$ energies  
 for elements between $6 \leq Z \leq 30$. The red triangles and green circles
show the difference with NIST for the updated Cloudy (C17.03) energies and the old Cloudy
energies appearing in C17.02 and before, respectively. The figure also shows the absolute values of the energy accuracy of current and future X-ray observatories.
The solid and 
dashed magenta lines represent the energy accuracy 
of Chandra HEG and MEG, 
respectively\footnote{\url{https://cxc.harvard.edu/proposer/POG/html/chap8.html}}.
The solid and dotted brown lines indicate the accuracy of RGS1 (1$^{\rm st}$ and 2$^{\rm nd}$ order) and RGS2 (1$^{\rm st}$ and 2$^{\rm nd}$ order) onboard XMM-Newton\footnote{\url{https://xmm-tools.cosmos.esa.int/external/xmm_user_support/documentation/uhb/rgs.html}}.
The solid black and blue lines represent the energy accuracy of XRISM \citep{2018JLTP..193..991I} and Athena \citep{2016SPIE.9905E..2FB}, respectively. The updated K$\alpha$ energies in the revised Cloudy
release (C17.03) are $\sim$ 15-4000 times more precise than that of C17.02. This energy precision is also much superior to the energy accuracy of the current and future X-ray instruments. 
The improved Cloudy energies will therefore be in  excellent agreement
with the future microcalorimeter observations. 

Finally, we created a patch file: `H\_total\_correction.diff', which includes the changes
leading up to the updated K$\alpha$ energies ($E_{\rm K\alpha}^{\rm new}$).  
These changes
will be part of C17.03, and are now posted to the Cloudy user group\footnote{\url{https://cloudyastrophysics.groups.io/g/Main/topics}}.
Updates on the development of Cloudy are posted to its wiki\footnote{\url{https://trac.nublado.org/wiki/NewC17}}.

The improvement described here  is made only for the
K$\alpha$ energies for elements between Carbon (Z=6) and Zinc (Z=30) since it is negligible
for lighter elements.
The doublet splitting for the 2p levels is not included in this update but is part of the thesis work and will
be incorporated into the next major  release of Cloudy. 
This future version will read extensive data files from NIST instead of using the above correction.



\acknowledgments
We acknowledge support by NSF (1816537, 1910687), NASA (17-ATP17-0141, 19-ATP19-0188), and STScI (HST-AR-15018).
MC also acknowledges support from STScI (HST-AR-14556.001-A). SB acknowledges financial support from the Italian Space Agency (grant 2017-12-H.0) and from the PRIN MIUR project ‘Black Hole winds and the BaryonLife Cycle of Galaxies: the stone-guest at the galaxy evolution supper’, contract 2017-PH3WAT.

\bibliography{FeKaWavelengths}{}
\bibliographystyle{aasjournal}

\end{document}